\documentclass[reprint,
superscriptaddress,
showpacs,
amsmath,amssymb,
aps,
pra,
]{revtex4-1}

\usepackage{amsmath}
\usepackage{amssymb}
\usepackage{graphicx}
\usepackage{bm}
\usepackage{hyperref}
\usepackage{enumerate}
\usepackage{color}

\begin{document}

\preprint{APS/Non-harmonic}

\title{Resonance fluorescence of a laser cooled atom in a non-harmonic potential}

\author{Ralf Betzholz}
\email{ralf.betzholz@physik.uni-saarland.de}
\affiliation{Theoretische Physik, Universit\"at des Saarlandes, D-66123 Saarbr\"ucken, Germany}
\author{Marc Bienert}
\affiliation{Theoretische Physik, Universit\"at des Saarlandes, D-66123 Saarbr\"ucken, Germany}
\affiliation{Hohenzollern Gymnasium, D-72488 Sigmaringen, Germany}

\date{\today}

%\pacs{37.10.Vz,  %Mechanical effects of light on atoms, molecules, and ions
%	      37.10.-x, %Atom, molecule, and ion cooling methods}
%	     32.50.+d, %Fluorescence, phosphorescence (including quenching)
%} 

\begin{abstract}
We investigate a single laser driven atom trapped in a non-harmonic potential. We present the performance of ground-state laser cooling and Doppler cooling and the signatures of the center-of-mass motion in the power spectrum of the scattered light. In order to illustrate the results we provide two explicit examples for the confining potential: the infinite square well and the Morse potential.
\end{abstract}
 
\maketitle

\section{Introduction}
\label{sec:intro}

In 1953, Dicke investigated the effect of photon recoil on the spectrum of light scattered by an atom~\cite{dicke1953}: When the atom emits a single photon, it experiences a recoil and thereby changes its motional state. The required energy for this acceleration is taken the photon, which shows a shifted frequency after the scattering event. In order to describe this effect most clearly, Dicke presented its calculation based on a simple confining potential, namely an infinite square well.

Trapping atoms in confining potentials has meanwhile become state of the art in experiments and exploiting
the momentum recoil due to the spontaneous emission of single photons is routinely applied in laser cooling in today's laboratories\cite{cohen1998,phillips1998,chu1998,metcalf2001}. Laser cooling is achieved when the laser parameters are chosen such that photon scattering processes that diminish the atomic motional energy prevail over transitions that heat the motion~\cite{wineland1979,dalibard1985}. For trapped ions or ground-state cooled atoms, it is justified to approximate the trapping potential harmonically. In the Lamb-Dicke limit~\cite{javanainen1981} the atomic wavepacket is spatially confined on a scale much smaller than the laser wavelength. Laser cooling theory then predicts that light scattering drives the atomic motion towards a thermal state~\cite{lindberg1984b}. In the spectrum of resonance fluorescence~\cite{raab2000} two distinct peaks emerge, the motional sidebands, as the Stokes- and anti-Stokes components of the scattered light.

In this article we 
%revisit the primary work of Dicke and 
investigate the light scattering at an atom trapped in a non-harmonic potential in the Lamb-Dicke regime. By this analysis we extend the theoretical tools for the description of laser cooling~\cite{lindberg1984a,cirac1992}  and the corresponding spectrum of resonance fluorescence~\cite{cirac1993,lindberg1986,bienert2004} to the case of non-harmonic potentials. Compared to the previous works, the details of the theoretical treatment for arbitrarily shaped potentials are presented and compared to the well-known harmonic case. This extension becomes relevant, for example, for atoms cooled in optical lattices~\cite{gatzke1997,hamann1998,bloch2008}, where the harmonic approximation can be insufficient, especially at higher temperatures. Similarly, in cavity cooling experiments~\cite{horak1997,pinkse2000,maunz2004,ritsch2013} the trapping potential's anharmonicity can manifest itself in the cooling dynamics and the spectrum of scattered light.

The work presented here focuses on the expansion of the theoretical description, guided by Dicke's original work based on a particle in a box. The infinite square, but also the Morse potential, is used to exemplify the application of the extended theory. Both potentials support an analytic solution of the eigenvalue problem given by the time independent Schr\"odinger equation for the center-of-mass motion, allowing for a clear and comprehensive treatment of laser cooling and the analysis of the scattered light. The Morse potential describes the dynamics of the relative coordinate of diatomic molecules~\cite{morse1929-1,morse1929-2} and therefore is of essential interest for the cooling~\cite{morigi2007,nguyen2011} and spectroscopy~\cite{roy2013} of such systems.

The presented approach allows to identify details in the perturbative description in the Lamb-Dicke regime that are connected with the non-degeneracy of transition frequencies between the motional eigenstates. For both potentials the resulting steady state of the center-of-mass motion cannot be written in terms of a thermal distribution and the motional sidebands consist of a series of peaks of finite width whose spectral position is connected to the transition frequencies between the relevant vibrational states. 

The article is organized as follows: In Sec.~\ref{sec:1} we introduce the system and present in Sec.~\ref{sec:2} the elements of the theory of light scattering that are necessary to describe the signatures of the atomic motion in the spectrum of resonance fluorescence. In order to discuss the results and to illustrate the deviations from harmonic trapping potentials, we present in Sec.~\ref{sec:potentials} the laser cooling performance and the motional sidebands of the scattered light for the two previously mentioned potentials. Finally, in Sec.~\ref{sec:conclusion} we draw the conclusions.

\section{System}
\label{sec:1}
\begin{figure}[ht]
\begin{center}
\includegraphics[width=0.7\linewidth]{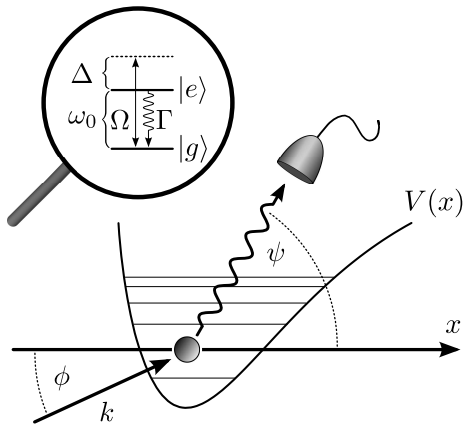}
\caption{\label{fig:system}Two-level atom with excited state linewidth $\Gamma$. The transition is detuned from an incidence laser with wave number $k$ by $\Delta$. The center of mass is confined along the $x$-direction in the non-harmonic potential $V(x)$. A detector is placed in the far field at an angle $\psi$ while the laser illuminates the atom under the angle $\phi$ with respect to the motional axis.}
\end{center}
\end{figure}

We investigate the radiation scattered by a single laser-cooled two-level atom. Along the $x$-direction the atom's center of mass is tightly trapped in a non-harmonic potential $V(x)$ and we restrict ourselves to the one-dimensional problem. The relevant electronic states are the ground state $\vert g\rangle$ and the excited state $\vert e\rangle$ which are energetically separated by the transition frequency $\omega_0$. A running wave laser with wave number $k$ irradiates the atom under an angle $\phi$ with respect to the motional axis and drives the dipole transition between the two levels with Rabi frequency $\Omega$. The laser with frequency $\omega_\mathrm{L}$ is detuned from the atomic transition by $\Delta=\omega_\mathrm{L}-\omega_0$. The scattered photons are recorded by a narrow-band detector positioned at an angle $\psi$ from the axis of motion. The rate of spontaneous emission of the two-level system is given by $\Gamma$. Figure~\ref{fig:system} depicts a sketch of the setup.

We assume that the coupling of the light field to the atomic motion is in a regime where the size of the center-of-mass wave packet is much smaller than the laser wavelength. Formally this can be expressed by the necessary condition
\begin{align}
\label{eq:lamb-dicke}
\eta=k\xi\ll 1,
\end{align}
 for the smallness parameter $\eta$, where $\xi=\sqrt{\langle x^2\rangle_0-\langle x\rangle_0^2}$ denotes the position uncertainty of the atomic ground-state wave function in the potential $V(x)$. The requirement on the atomic localization puts constraints on the occupation of higher excited motional states: Only a sufficiently low mean occupation number $\bar{m}$ of energy eigenstates in the trapping potential is allowed for the treatment presented in this work to be valid.
We note that for harmonic trapping $\eta$ corresponds to the Lamb-Dicke parameter~\cite{javanainen1981,lindberg1984a,stenholm1986}.

The Hamiltonian of the system is composed of an internal part $H_{\rm I}$ describing the electronic states, an external part $H_{\rm E}$ accounting for the atomic center-of-mass motion and the coupling $W(x)$ of those two degrees of freedom by the mechanical effects of the laser light, viz.
\begin{align}
\label{eq:hamiltonian1}
H=H_{{\rm I}}+H_{{\rm E}}+W(x).
\end{align}
In the frame rotating with the laser frequency the internal Hamiltonian is given by
\begin{align}
\label{eq:hamiltonian_int}
H_{{\rm I}}=-\hbar\Delta\sigma_+\sigma_-,
\end{align}
where $\sigma_+=\vert e\rangle\langle g\vert$ and $\sigma_-=\vert g\rangle\langle e\vert$ represent the atomic raising and lowering operators, respectively. The external Hamiltonian reads
\begin{align}
\label{eq:hamiltonian_ext}
H_{{\rm E}}=\frac{p^2}{2M}+V(x)
\end{align}
with the atomic mass $M$ and the momentum operator $p$. The coupling of the electronic and motional degrees of freedom due to the laser field takes on the form
\begin{align}
\label{eq:hamiltonian_int2}
W=\hbar\frac{\Omega}{2}\left(\sigma_+ e^{ikx\cos\phi}+{\rm H.c.}\right).
\end{align}
To complete the description, we take electronic relaxation processes into account using a master-equation
formalism. The time evolution of the system's density operator $\varrho$, covering the internal and external degrees of freedom, including spontaneous emission, is generated by the Liouville operator $\mathcal{L}$ and obeys
\begin{align}
\label{eq:master_equation1}
\frac{\partial\varrho}{\partial t}=&\mathcal{L}\varrho\nonumber\\
=&\frac{1}{i\hbar}[H,\varrho]+\frac{\Gamma}{2}\int_{-1}^1 du \, w(u) \mathcal{D}[\sigma_-e^{-ikxu}]\varrho
\end{align}
with the abbreviation $\mathcal{D}[X]\varrho=2X\varrho X^\dagger-X^\dagger X\varrho-\varrho X^\dagger X$ for superoperators of Lindblad form. The normalized radiation pattern $w(u)$ for the considered transition describes the probability of emitting a photon
at an angle $\psi=\arccos u$ while the exponential accounts for momentum recoils due to spontaneously emitted photons projected on the axis of motion. The specific form of  the symmetric function $w(u)$ depends on the details of the electronic transition~\cite{lindberg1984a}. 

\section{Theory of light scattering}
\label{sec:2}
Under stationary conditions the spectral signal at the detector is given by~\cite{lindberg1986,cirac1993}
\begin{align}
\label{eq:spectrum1}
\mathcal{S}(\omega) = {\rm Re}\int_{0}^{\infty} dt\, e^{-i(\omega-\omega_{\rm L}) t} \langle D_+(t)D_-(0)\rangle_{\rm st},
\end{align}
where in the far field the two mutually adjoint generalized atomic lowering and raising operators have the form 
\begin{align}
D_-(t)=&\sigma_-(t)e^{-ikx\cos\psi},\nonumber\\
D_+(t)=&\sigma_+(t)e^{ikx\cos\psi},\label{eq:Dpm}
\end{align}
respectively. The exponential term in Eq.~\eqref{eq:Dpm} accounts for the recoil of the photon of wave number $k$ spontaneously emitted along the direction specified by $\psi$, projected on the axis of motion. 

\subsection{Spectral decomposition and perturbative expansion of the Liouville operator}
\label{sec:spectral_decomp}
A convenient way to evaluate the power spectrum~\eqref{eq:spectrum1} is to employ the spectral decomposition of the Liouville operator defined in Eq.~\eqref{eq:master_equation1}, the so-called damping basis~\cite{briegel1993,barnett2000}. This method has already been applied for the description of light scattering~\cite{cirac1993,bienert2004,torres2007,torres2011} and laser cooling~\cite{lindberg1984a,cirac1992} of harmonically trapped atoms in the Lamb-Dicke limit.

Formally the solution of the master equation~\eqref{eq:master_equation1} can be achieved by solving the eigenvalue equations of $\mathcal{L}$ for the left and right eigenelements which read 
\begin{align}
\label{eq:eigenvalueequation}
\mathcal{L}\hat{\varrho}_\lambda&=\lambda\hat{\varrho}_\lambda,\\
\label{eq:eigenvalueequationleft}
\check{\varrho}_\lambda^\dagger\mathcal{L}&=\lambda\check{\varrho}_\lambda^\dagger.
\end{align}
The eigenelements are orthogonal with respect to the scalar product ${\rm Tr}\{\check{\varrho}_\lambda^\dagger\hat{\varrho}_{\lambda'}\}=\delta_{\lambda,\lambda'}$ and we assume they for  a complete set, formally expressed in
\begin{align}
\label{eq:closure}
\sum_{\lambda}\hat{\varrho}_\lambda\otimes\check{\varrho}_\lambda=1,
\end{align}
where the action of the projectors on an arbitrary operator $X$ is given by $(\hat{\varrho}_\lambda\otimes\check{\varrho}_\lambda) X={\rm Tr}\{\check{\varrho}_\lambda^\dagger X\} \hat{\varrho}_\lambda$. 

A small value of $\eta$ suggests an expansion of the Liouville operator $\mathcal{L}$ with techniques described in \cite{torres2007,bienert2004}. Up to second order in $\eta$ we write the Liouville operator as $\mathcal{L}=\mathcal{L}_0+\mathcal{L}_1+\mathcal{L}_2$, where the subscript indicates the order. The individual terms read
\begin{align}
\mathcal{L}_0\varrho&=\mathcal{L}_{\rm I}\varrho+\mathcal{L}_{\rm E}\varrho\nonumber\\
\label{eq:liouvillians0}
&=\frac{1}{i\hbar}[H_{{\rm I}}+W_0,\varrho]+\frac{\Gamma}{2}\mathcal{D}[\sigma_-]\varrho+\frac{1}{i\hbar}[H_{{\rm E}},\varrho],\\
\label{eq:liouvillians1}
\mathcal{L}_1\varrho&=\frac{1}{i\hbar}[W_1x,\varrho],\\
\label{eq:liouvillians2}
\mathcal{L}_2\varrho&=\frac{1}{i\hbar}[W_2x^2,\varrho]+\alpha\frac{\Gamma}{2}k^2\sigma_-\mathcal{D}[x]\varrho\sigma_+
\end{align}
with the definition $\alpha=\int_{-1}^1 du \, w(u)u^2$ (which evaluates to $2/5$ for the dipole pattern used here~\cite{lindberg1984a}). The expansion 
\begin{align}
\label{eq:coupling_expansion}
W_n=\frac{1}{n!}\left.\frac{\partial^nW}{\partial x^n}\right|_{x=0}
\end{align}
of the interaction Hamiltonian up to second order is explicitly given by
\begin{align}
\label{eq:coupling_expansion_exlicit}
W_0&=\hbar\frac{\Omega}{2}\left(\sigma_++\sigma_-\right),\\ 
W_1&=i\hbar\frac{\Omega}{2} k\cos\phi\left(\sigma_+-\sigma_-\right),\\
W_2&=-\hbar\frac{\Omega}{4}k^2\cos^2\phi\left(\sigma_++\sigma_-\right).
\end{align}
We solve the eigenvalue equations~\eqref{eq:eigenvalueequation} and~\eqref{eq:eigenvalueequationleft} in zeroth order of $\eta$ and then perform perturbation theory to obtain the eigenvalues and eigenelements in higher orders. 

The zeroth order Liouville operator Eq.~\eqref{eq:liouvillians0} does not couple the internal and external degrees of freedom. Hence, the eigenelements 
\begin{align}
\label{eq:factorize}
\hat{\varrho}_\lambda^{(0)}&=\hat{\rho}_{\lambda_{\rm I}}\hat{\mu}_{\lambda_{\rm E}},\\
\check{\varrho}^{\dagger (0)}_\lambda&=\check{\rho}_{\lambda_{\rm I}}^\dagger\check{\mu}_{\lambda_{\rm E}}^\dagger
\end{align}
factorize, where $\rho$ and $\mu$ denote eigenelements of the internal and external degrees of freedom, respectively. The eigenvalues are $\lambda_0=\lambda_{\rm I}+\lambda_{\rm E}$, with $\lambda_{\rm I}$ and $\lambda_{\rm E}$ denoting the internal and external eigenvalues of $\mathcal{L}_{\rm I}$ and $\mathcal{L}_{\rm E}$, respectively. Therefore, we only have to solve the eigenvalue equations of the internal and external motion separately, which read 
\begin{align}
\label{eq:eigenvalue_equation1}
\mathcal{L}_{\rm I}\hat{\rho}_{\lambda_{\rm I}}=\lambda_{\rm I}\hat{\rho}_{\lambda_{\rm I}},&\quad	\check{\rho}_{\lambda_{\rm I}}^{\dagger}\mathcal{L}_{\rm I}=\lambda_{\rm I}\check{\rho}_{\lambda_{\rm I}}^{\dagger},\\
\label{eq:eigenvalue_equation3}
\mathcal{L}_{\rm E}\hat{\mu}_{\lambda_{\rm E}}=\lambda_{\rm E}\hat{\mu}_{\lambda_{\rm E}},&\quad	\check{\mu}_{\lambda_{\rm E}}^{\dagger}\mathcal{L}_{\rm E}=\lambda_{\rm E}\check{\mu}_{\lambda_{\rm E}}^{\dagger}.
\end{align}

The eigenvalue equations for the internal Liouville operator can be readily solved using a matrix representation of the superoperator~\cite{jakob2003}. In App.~\ref{app:internal} we give explicit expressions including the steady state $\rho_{\rm st}$ of the internal dynamics.

The external Liouvillian~\eqref{eq:liouvillians0} does not include any non-unitary terms and its eigenelements
\begin{align}
\label{eq:ext_eigenelements}
\hat{\mu}_{nm}&=\vert n\rangle\langle m\vert,\\
\check{\mu}_{nm}^\dagger&=\vert m\rangle\langle n\vert
\end{align}
can be constructed from the energy eigenstates $|n\rangle$ satisfying
\begin{align}
\label{eq:ext_energystates}
H_{\rm E}\vert n\rangle=\varepsilon_n\vert n\rangle.
\end{align}
The corresponding external eigenvalues $\lambda_{nm}=i\omega_{nm}$ contain the transition frequencies 
\begin{align}
\label{eq:transition_freq}
\omega_{nm}=\frac{\varepsilon_m-\varepsilon_n}{\hbar}
\end{align}
between the energy eigenstates $|m\rangle$ and $|n\rangle$.

The perturbative corrections of interest for later calculations are the ones of first order,
\begin{align}
\label{eq:first_order_correction1}
\check{\varrho}_{\lambda}^{\dagger(1)}=&\check{\varrho}_{\lambda}^{\dagger(0)}\mathcal{L}_1\left(\lambda_0-\mathcal{L}_0\right)^{-1}\mathcal{Q}_\lambda,\\
\label{eq:first_order_correction2}
\hat\varrho_{\lambda}^{(1)}=&\left(\lambda_0-\mathcal{L}_0\right)^{-1}\mathcal{Q}_\lambda\mathcal{L}_1\hat\varrho_{\lambda}^{(0)}.
\end{align}
Here subscripts of the eigenvalues and superscripts of the eigenelements again label the corresponding order of $\eta$. The projectors introduced in Eqs.~\eqref{eq:first_order_correction1} and~\eqref{eq:first_order_correction2} are given by $\mathcal{Q}_\lambda=1-\mathcal{P}_\lambda$ and  $\mathcal{P}_\lambda=\hat\varrho_{\lambda}^{(0)}\otimes\check{\varrho}_{\lambda}^{\dagger(0)}$.

\subsection{Resonance fluorescence}
The time evolution of the operators in expression~\eqref{eq:spectrum1} for the spectrum of resonance fluorescence is determined
by the Liouville operator $\mathcal{L}$ and can be calculated using the quantum regression theorem~\cite{lax1963,carmichael2002}. Together with the 
eigenvalue equations~\eqref{eq:eigenvalueequation} and~\eqref{eq:eigenvalueequationleft} of the Liouville operator and the completeness relation~\eqref{eq:closure} the spectrum formula~\eqref{eq:spectrum1} can be cast into the form
\begin{align}
\label{eq:spectrum3}
\mathcal{S}(\omega) = {\rm Re} \sum\limits_{\lambda}  \frac{w_{\lambda}}{i(\omega-\omega_{\rm L}) - \lambda },
\end{align}
i.e. we can decompose the spectrum into contributions connected to the eigenvalues of the Liouville operator, weighted by
\begin{align}
\label{eq:weight_factors1}
w_\lambda={\rm Tr}\{D_+\hat{\varrho}_\lambda\}{\rm Tr}\{\check{\varrho}^\dagger_\lambda D_-\varrho_{\rm st}\}.
\end{align}
Depending on the real and imaginary parts of $w_\lambda$ the spectrum consists of a superposition of Lorentzians and Fano profiles.

We are mainly interested in the signatures of the atomic motion in the spectrum of the scattered light. Therefore we only focus on contributions fulfilling the following criteria: 
(i) We only take the first non-vanishing correction, i.e. the second order in $\eta$, of the spectrum into account. (ii) We only consider eigenvalues with $\lambda_{\rm I}=0$ giving the motional sidebands of the elastic peak~\footnote{For the harmonic trapping potential the motional sidebands of the inelastic peaks were reported in Ref.~\cite{bienert2006}.}. (iii) We do not report the contribution $\lambda_{\rm I}=0$ and $\lambda_{\rm E}=0$ resulting in a correction to the Rayleigh peak.

In order to evaluate the factors~\eqref{eq:weight_factors1} we expand the generalized atomic lowering operators as $D_-=D_-^{(0)}+D_-^{(1)}+...$ with
\begin{align}
\label{eq:gen_dipole1}
D_-^{(0)}&=\sigma_-,\\
\label{eq:gen_dipole2}
D_-^{(1)}&=-ik\cos\psi\,\sigma_-x
\end{align}
and likewise the weight factors according to $w_\lambda=w_\lambda^{(0)}+w_\lambda^{(1)}+...$ in orders of $\eta$. In this expansion the zeroth order weight factors give rise to the Mollow-type spectrum of a laser driven two-level system~\cite{mollow1969}. It turns out that the first order does not contribute, while the second order takes on the form
\begin{align}
\label{eq:weight_factors2}
w^{(2)}_\lambda=\sum_{\substack{\alpha+\beta+\gamma\\+\delta+\epsilon=2}}{\rm Tr}\big\{D_+^{(\alpha)}\hat{\varrho}_\lambda^{(\beta)}\big\}{\rm Tr}\big\{\check{\varrho}_\lambda^{\dagger (\gamma)}D_-^{(\delta)}\varrho_{\rm st}^{(\epsilon)}\big\}.
\end{align}
The only contributions to the weight factors that fulfill the conditions (i)-(iii) read
\begin{align}
\label{eq:weight_factors3}
w&_\lambda^{(2)}={\rm Tr}\big\{D_-^{(0)}\hat{\varrho}^{(1)}_\lambda+D_-^{(1)}\hat{\varrho}^{(0)}_\lambda\big\}\times\nonumber\\
&{\rm Tr}\Big\{\big[\check{\varrho}^{\dagger(1)}_\lambda D_+^{(0)}+\check{\varrho}^{\dagger(0)}_{\lambda}D_+^{(1)}\big]\varrho^{(0)}_{\rm st}+\check{\varrho}^{\dagger(0)}_{\lambda}D_+^{(0)}\varrho^{(1)}_{\rm st}\Big\}
\end{align}
where in $\varrho_{\rm st}^{(0)}=\rho_{\rm st}\mu_{\rm st}$ the external steady state has the form
\begin{align}
\label{eq:ext_steady}
\mu_{\rm st}=\sum_j p_j \vert j\rangle\langle j\vert
\end{align}
in the energy eigenbasis~\eqref{eq:ext_energystates} of the external Hamiltonian. The populations $p_j=\text{Tr}_{\text{I}}\langle j \vert\rho\vert j \rangle$ in the energy eigenstates $\vert j\rangle$ are determined by laser cooling as discussed in the next section.

An outline of the evaluation of the factors~\eqref{eq:weight_factors3} is presented in App.~\ref{app:sidebands}. The spectrum of resonance fluorescence can be brought into the form
\begin{align}
\label{eq:spectrum4}
\mathcal{S}_{\rm sb}(\omega&) = {\rm Re}\sum_{n\neq m}\frac{\vert \langle n\vert x\vert m\rangle\vert^2}{i(\omega-\omega_{\rm L}-\omega_{nm})-\lambda_{nm}^{(1)}-\lambda_{nm}^{(2)}}\nonumber\\
&\times\left[p_m |\tilde{r}(\omega_{nm})|^2+(p_n-p_m)\tilde{r}(\omega_{nm})q(\omega_{nm})\right]
\end{align}
where $\tilde{r}(\omega)=r(\omega)-[\Delta+i\Gamma/2]\Omega k\cos\psi/2N$ and $N=\Gamma^2/4+\Delta^2+\Omega^2/2$. Furthermore, we defined the two functions~\cite{cirac1993}
\begin{align}
\label{eq:r}
r(\omega) & =\frac{1}{\hbar}\int_0^{\infty} dt \, {\rm e}^{-i\omega t} \, \langle [\sigma_+(t), W_1(0)]  \rangle_{{\rm st}},\\
\label{eq:q}
q(\omega) & = \frac{1}{\hbar}\int_{-\infty}^{\infty} dt \, {\rm e}^{-i\omega t} \, \langle W_1(t) \sigma_-(0) \rangle_{{\rm st}}.
\end{align}
Explicit expressions for $r(\omega)$ and $q(\omega)$ are given in App.~\ref{app:r_and_q}. In the denominator of Eq.~\eqref{eq:spectrum4} we used the perturbative expansion $\lambda_{nm}=\lambda_{nm}^{(0)}+\lambda_{nm}^{(1)}+\lambda_{nm}^{(2)}$ of the eigenvalues. The higher order contributions of this expansion introduce a finite linewidth of the motional sidebands, since the denominator of Eq.~\eqref{eq:spectrum4} is purely imaginary in zeroth order.

In the well studied case of a harmonic trapping potential the external eigenvalues $\lambda_{nm}$ are degenerate due to the equidistant eigenenergies of the potential. In that specific case, perturbation theory for degenerate eigenvalues has to be performed. In the problem treated here the potential shows an appreciable anharmonicity such that generally only the eigenvalue $\lambda_{\rm E}=0$ is degenerate. Hence, perturbation theory for non-degenerate eigenvalues has to be applied. Such an approach is valid if the splitting of the eigenvalues due to the interaction with the laser is small compared to the energy differences of the vibrational levels~\cite{morigi2003}. Explicitly the condition 
\begin{align}
\eta\Omega\ll{\rm min}_{n,n'\ne n}\vert\omega_{nn'}\vert
\end{align}
has to be fulfilled.

%which for the potentials examined in a later section is given in the parameter regime we consider.

The first order correction to the eigenvalues is given by
\begin{align}
\label{eq:eigenvals1}	
\lambda_1={\rm Tr}\big\{\check{\varrho}^{\dagger(0)}_\lambda\mathcal{L}_1\hat{\varrho}^{(0)}_\lambda\big\}
\end{align}
which for $\lambda_0=\lambda_{nm}$ can be written as $\lambda^{(1)}_{nm}=i\delta\omega^{(1)}_{nm}$ with
\begin{align}
\label{eq:eigenvals2}	
\delta\omega^{(1)}_{nm}=\frac{\Gamma\Omega^2 k\cos\phi}{4N}\big(\langle n\vert x\vert n\rangle-\langle m\vert x\vert m\rangle\big).
\end{align}
This constitutes a shift of the peak positions but does not add a finite width. We further note that this shift vanishes in all even potentials due to the parity of the eigenstates. The second order correction
\begin{align}
\label{eq:eigenvals3}	
\lambda_2={\rm Tr}\big\{\check{\varrho}^{\dagger(0)}_\lambda\left[\mathcal{L}_2+\mathcal{L}_1(\lambda_0-\mathcal{L}_0)^{-1}\mathcal{Q}_\lambda\mathcal{L}_1\right]\hat{\varrho}^{(0)}_\lambda\big\}
\end{align}
shows a non-vanishing real part. Their evaluation is sketched in App.~\ref{app:sidebands_width} where it is shown that for $\lambda_0=\lambda_{nm}$ one obtains $\lambda^{(2)}_{nm}=i\delta\omega^{(2)}_{nm}-\gamma_{nm}$ with the second order frequency shift
\begin{align}
\delta\omega&^{(2)}_{nm}=\frac{\Delta\Omega^2k^2\cos^2\phi}{4N} \left(\langle n\vert x^2\vert n\rangle-\langle m\vert x^2\vert m\rangle\right)\nonumber\\
&-{\rm Im}\sum_j\left[s(\omega_{jn})\vert\langle j\vert x\vert n\rangle\vert^2+s(\omega_{jm})\vert\langle j\vert x\vert m\rangle\vert^2\right]
\end{align}
and the real part
\begin{align}
\label{eq:linewidths}
\gamma_{nm}=\frac{1}{2}\sum_j\left(A_{jn}+A_{jm}\right)-D\langle n\vert x\vert n\rangle \langle m\vert x\vert m\rangle
\end{align}
describing finite width of the sidebands. In this expression we introduced the diffusion coefficient 
\begin{align}
D=\alpha\Gamma k^2{\rm Tr}\{\sigma_+\sigma_-\rho_{\rm st}\}=\frac{\Gamma\Omega^2k^2}{10N}
\end{align}
and the transition rates
\begin{align}
\label{eq:matrix_coefficients1}
A_{nm} &=\left[2\,{\rm Re}\, s(\omega_{nm})+D\right] \vert\langle n\vert x\vert m\rangle\vert^2
\end{align} 
 with the fluctuation spectrum
\begin{align}
\label{eq:s_omega}
s(\omega)=\frac{1}{\hbar^2}\int_0^{\infty}dt\, e^{i\omega t}\langle W_1(t)W_1(0)\rangle_{\rm st}.
\end{align}
Here, the two-time correlation function $s(\omega)$ (for explicit expressions see App.~\ref{app:r_and_q}) is evaluated in the steady state $\rho_{\rm st}$,  viz. $\langle X\rangle_{\rm st}={\rm Tr}\{X\rho_{\rm st}\}$. In the next section we will see that the coefficients $A_{nm}$ indeed describe the rate of population transfer between the energy eigenstate $\vert m\rangle$ and $\vert n\rangle$ due to the mechanical effects of light scattering. Inspecting $\gamma_{nm}$ defined in Eq.~\eqref{eq:linewidths} we find that the width of a sideband peak that originates in a transition from $\vert m\rangle$ to $\vert n\rangle$ involves a sum over the rates of transitions from $\vert m\rangle$ and $\vert n\rangle$ to all other states $|j\rangle$, i.e. $A_{jm}$ and $A_{jn}$. We note that since ${\rm Re}\,s(\omega)>0$ one also finds $\gamma_{nm}>0$.

\subsection{Cooling of the atomic motion}
In zeroth order of $\eta$ the steady state $\mu_{\rm st}$ cannot be determined uniquely since the eigenvalue $\lambda_{\rm E}=0$ is infinitely degenerate. This degeneracy is lifted in second order perturbation theory and the unique steady state of laser cooling can be found by adiabatic elimination of the internal degrees of freedom~\cite{javanainen1984,lindberg1984a,stenholm1986,gardiner2004}. This procedure applied to Eq.~\eqref{eq:master_equation1} yields the equation 
\begin{align}
\label{eq:reduced_dynamics}
\mathcal{P}_0\left(\mathcal{L}_1\mathcal{L}_0^{-1}\mathcal{Q}_0\mathcal{L}_1-\mathcal{L}_2\right)\mathcal{P}_0\rho_{\rm st}\mu_{\rm st}=0
\end{align}
for the external steady state $\mu_{\rm st}$. Using the representation~\eqref{eq:ext_steady} of $\mu_{\rm st}$ and performing a partial trace ${\rm Tr_{\rm I}}\{\cdot\}$ over the internal degrees of freedom in Eq.~\eqref{eq:reduced_dynamics} results in the a recursive equation for the probabilities $p_j$. One can rewrite this equation as
\begin{align}
\label{eq:rateequation}
\sum_mA_{nm}p_m-\sum_mA_{mn}p_n=0
\end{align}
with $A_{nm}$ defined in Eq.~\eqref{eq:matrix_coefficients1}, see App.~\ref{app:external_steady}. This set of equations determines the steady state of the atomic motion. In this rate equation the part of the coefficients $A_{nm}$ including $s(\omega)$ reflects the rate of transitions induced by the laser field while the diffusive part connected with the diffusion coefficient $D$ stems from spontaneous emission. 

\subsection{Comparison to harmonic trapping potential}
\label{sec:harmonic_comparison}
We conclude this section by a comparison with the harmonic trapping potential $V(x)=M\nu^2x^2/2$~\cite{javanainen1981,lindberg1984a,stenholm1986,cirac1992}. The matrix elements of the position operator are only non-zero between neighboring energy states, $\langle n\vert x\vert m\rangle\propto\sqrt{m} \delta_{n,m-1}+\sqrt{m+1}\delta_{n,m+1}$, allowing only transitions between adjacent vibrational levels. This is directly reflected in the transition rates $A_{nm}$ which are also non-zero only for $n=m\pm1$, resulting in a recurrence relation Eq.~\eqref{eq:rateequation} that has the form of the detailed balance condition
\begin{align}
\label{eq:detailed_balance}
nA_+p_{n-1}+(n+1)A_-p_{n+1}=[nA_-+(n+1)A_+]p_n
\end{align}
with $A_\pm=2\,{\rm Re}\,s(\mp\nu)+D$~\cite{cirac1992}. The normalized solution is a thermal distribution $p_n=\bar{m}^n/(\bar{m}+1)^{n+1}$ with the mean occupation number $\bar{m}=A_+/(A_--A_+)$. The steady state~\eqref{eq:ext_steady} of the atomic motion can be cast in the canonical form $\mu_{\rm st}=\exp(-\beta H_{\rm osc})/Z$ with the harmonic oscillator Hamiltonian $H_{\rm osc}$, the partition function $Z={\rm Tr}\{\exp(-\beta H_{\rm osc})\}$ and the inverse temperature $\beta$ implicitly defined via $\bar{m}=[\exp(\beta\hbar\nu)-1]^{-1}$. The harmonic potential is special in the sense that the effective dynamical equation~\eqref{eq:reduced_dynamics} takes the shape of a master equation of a harmonic oscillator in contact with a thermal reservoir~\cite{cirac1992}, resulting in a thermal distribution as a steady state. In contrast, non-harmonic potentials do {\it not} lead to a recursive relation as Eq.~\eqref{eq:detailed_balance} and generally do not lead to a thermal external state.

According to Eq.~\eqref{eq:spectrum4} the atomic motion causes the emergence of an infinite number of motional sidebands, each approximately centered around a possible transition between eigenstates of the external motion. The property that only transitions between neighboring levels can occur together with the degeneracy of the transition frequencies results in a sideband spectrum Eq.~\eqref{eq:spectrum5} consisting of {\it two} peaks centered around $\pm\nu$. The widths of both peaks can be obtained by employing perturbation theory for degenerate eigenvalues and is given by the cooling rate $\gamma=A_--A_+$, thereby giving a sideband spectrum of the form~\cite{cirac1993}
\begin{align}
	\label{eq:spectrum_ho}
	\mathcal{S}_{\rm sb}(\omega) =&{\rm Re}\frac{\bar{m} |\tilde{r}(\nu)|^2+\tilde{r}(\nu)q(\nu)}{i(\omega-\omega_{\rm L}-\tilde{\nu})^2+\gamma}\nonumber\\
	&+{\rm Re}\frac{(\bar{m}+1)|\tilde{r}(-\nu)|^2-\tilde{r}(-\nu)q(-\nu)}{i(\omega-\omega_{\rm L}+\tilde{\nu})^2+\gamma}
\end{align}
with a renormalized frequency $\tilde{\nu}$. %The smallness parameter for the harmonic potential is given by the well-known Lamb-Dicke parameter $\eta_{\rm osc}=k\sqrt{\hbar/2M\nu}$.

\section{Examples}
\label{sec:potentials}
We exemplify our results by means of two specific choices of the potential $V(x)$, namely the infinite square well and the Morse potential. For both potentials we will focus on two distinct parameter regimes of laser cooling: The regime of Doppler cooling ($\omega_{nm}\ll\Gamma$) and the regime of resolved sideband cooling ($\omega_{nm}\gg\Gamma$).

In the following we will assume the Rabi frequency to be small such that the atom is driven below saturation. In this case the function $s(\omega)$ in Eq.~\eqref{eq:s_omega} can be expanded as
\begin{align}
s(\omega)=\frac{\Omega^2}{4}k^2\cos^2\phi\,\frac{\Gamma/2+i(\omega+\Delta)}{\Gamma^2/4+(\omega+\Delta)^2}+\mathcal{O}(\Omega^4).
\end{align}
In the spectrum formula~\eqref{eq:spectrum4} we find $|\tilde{r}(\omega)|^2=\mathcal{O}(\Omega^2)$ while $\tilde{r}(\omega)q(\omega)=\mathcal{O}(\Omega^4)$. In the low intensity limit $\Omega\ll\Gamma$ this implies that we can write the motional sidebands of the elastic peak in second order of $\eta$ as
\begin{align}
\label{eq:spectrum5}
\mathcal{S}_{\rm sb}(\omega) = \sum_{n\neq m}\frac{\gamma_{nm}p_m\vert \langle n\vert x\vert m\rangle\vert^2|\tilde{r}(\omega_{nm})|^2}{\gamma_{nm}^2+(\omega-\omega_{\rm L}-\tilde{\omega}_{nm})^2}
\end{align}
where we introduced the renormalized frequencies $\tilde{\omega}_{n,m}=\omega_{n,m}+\delta\omega_{nm}^{(1)}+\delta\omega_{nm}^{(2)}$ that slightly shift the sideband peaks from the natural transitions between vibrational states.

\subsection{Infinite square well}
\label{subsec:well}
\begin{figure}[ht]
	\begin{center}
		\includegraphics[width=0.7\linewidth]{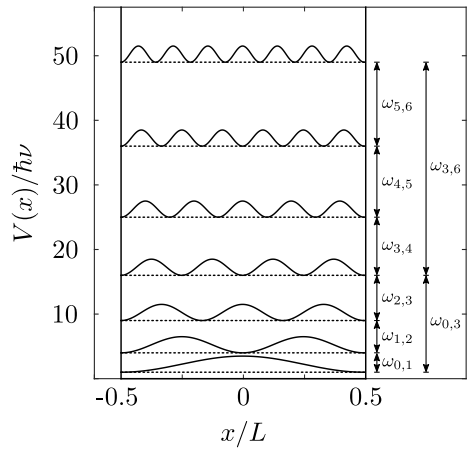}
		\caption{\label{fig:potential_well}Infinite square well potential with its first seven eigenenergies (dashed) and the corresponding probability densities $|\varphi_n(x)|^2$ (solid) with some transition frequencies of allowed transitions (in units of $\nu$).}
	\end{center}
\end{figure}

In Ref.~\cite{dicke1953} a study of light scattered by an atom confined in a one-dimensional square well (see Fig.~\ref{fig:potential_well}) of length $L$ has been performed. 
We revisit this problem using the theory developed in the previous sections. As a first example we therefore consider the atom to be trapped in the symmetric potential
\begin{align}
\label{eq:well_potential}
V(x)=\left\{
\begin{array}{cl}
\infty, & \ \text{for} \ |x|\ge L/2\\
0,& \ \text{for} \ |x|< L/2.
\end{array}
\right.
\end{align}
The eigensystem of $H_{\rm E}$ is given by
\begin{align}
\varepsilon_n=&(n+1)^2\hbar\nu,\\
\label{eq:states_well}
\varphi_n(x)=&
\left\{
\begin{array}{cl}
\sqrt{\frac{2}{L}}\sin((n+1)\pi x/L), \ \text{for\ odd} \ n\\
\sqrt{\frac{2}{L}}\cos((n+1)\pi x/L), \ \text{for\ even} \ n.
\end{array}
\right.
\end{align}
for $n=0,1,2,...$ with the frequency $\nu=\hbar\pi^2/2ML^2$. Figure~\ref{fig:potential_well} shows the potential well with the vibrational eigenenergies in units of $\hbar\nu$ and the probability density of the associated wavefunctions $\varphi_n(x)=\langle x\vert n\rangle$. The transition frequency between neighboring eigenstates increases linearly with $n$ according to 
\begin{align}
\omega_{n,n+1}=(2n+3)\nu.
\end{align}

With the help of the eigenstates~\eqref{eq:states_well} it is easy to calculate the matrix elements of the position operator which are needed to evaluate the coefficients $A_{nm}$ and the eigenvalue corrections Eqs.~\eqref{eq:eigenvals2} and~\eqref{eq:eigenvals3}. They are given by
\begin{align}
\langle n\vert x\vert m\rangle=-\frac{8(n+1)(m+1)(-1)^{(n+m+1)/2}L}{\pi^2[(n+1)^2-(m+1)^2]^2}
\end{align}
for $n,m$ of different parity and zero otherwise, being a consequence of the symmetry of the potential. The matrix elements of $x^2$ are also readily evaluated and their diagonal elements read
\begin{align}
\langle n\vert x^2\vert n\rangle=\frac{L^2}{12}\left[1-\frac{6}{(n+1)^2\pi^2}\right].
\end{align}
For the ground state $n=0$ this yields a smallness parameter, as introduced in Eq.~\eqref{eq:lamb-dicke}, of
\begin{align}
\eta=\frac{kL}{2\pi}\sqrt{\frac{\pi^2-6}{3}}\approx 0.18\,kL.
\end{align}
The parameter $kL$ was already mentioned in Ref.~\cite{dicke1953} to quantify the influence of the atomic motion on the emitted radiation.

The above mentioned matrix elements can now be used to determine the steady state of the atomic motion. Although the explicit shape of the rate equation~\eqref{eq:rateequation} is easily obtained it is nevertheless necessary to solve the rate equation numerically due to the complexity of the coefficient matrix  Eq.~\eqref{eq:rateequationmatrix}.

\subsubsection{Doppler regime}
In the Doppler regime the transition frequencies between the lower eigenstates are small compared to the linewidth of the atomic transition. To fulfill this criterion we choose $\nu=\Gamma/30$ which corresponds to a lowest transition frequency of $\omega_{0,1}=\Gamma/10$. 

\begin{figure}[h!]
\flushleft{(a)}\vspace{-4ex}
\begin{center}
\includegraphics[width=0.8\linewidth]{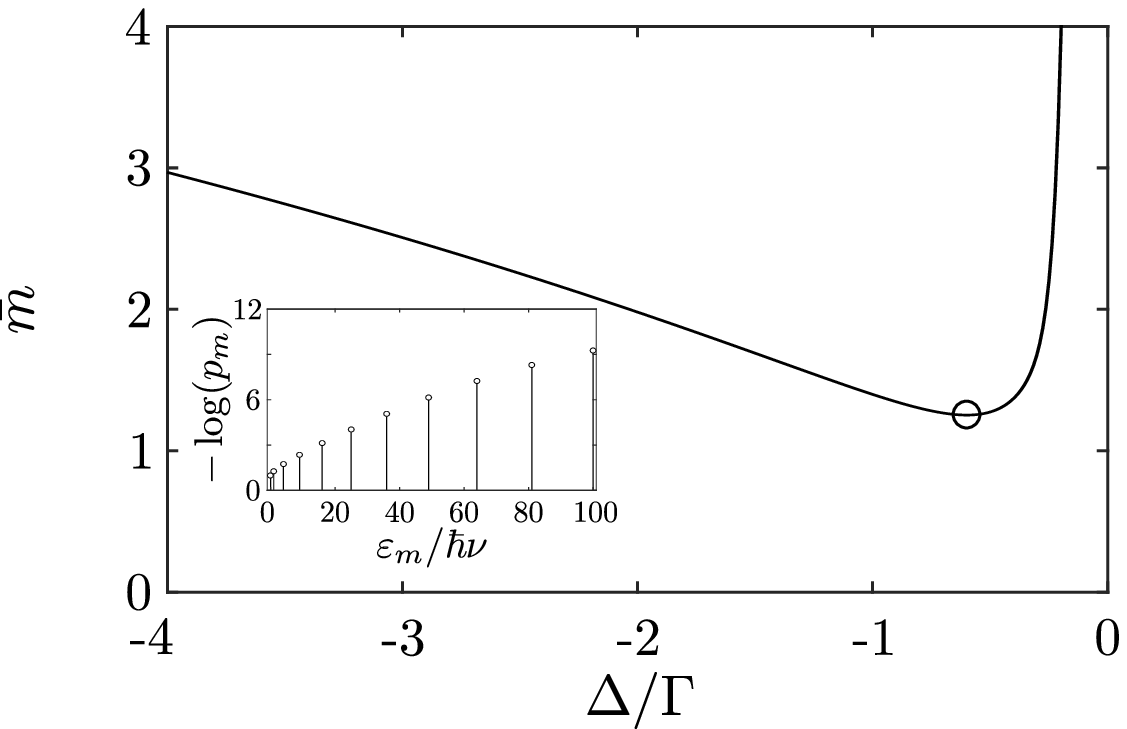}
\end{center}
\flushleft{(b)}\vspace{-4ex}
\begin{center}
	\includegraphics[width=0.8\linewidth]{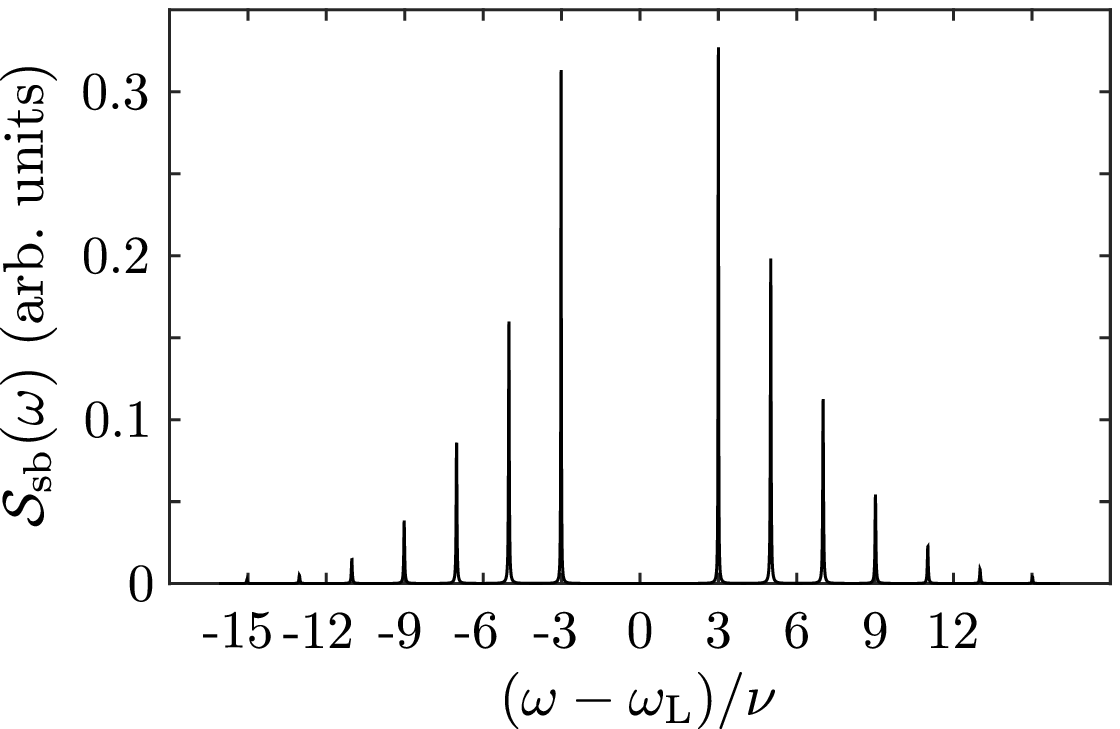}
\end{center}
\caption{\label{fig:well_doppler}Doppler cooling of an atom trapped in an infinite square well potential for $\nu=\Gamma/30$ ($\omega_{0,1}=\Gamma/10$), $\Omega=\Gamma/5$ and $\eta=0.1$. (a) Mean occupation $\bar{m}$ in dependence on the detuning $\Delta$. The circle indicates optimal cooling for $\Delta\approx-0.6\Gamma$ resulting in $\bar{m}\approx1.24$. The inset shows the logarithm of the steady state occupations $p_m$ ($m=0,...,10$) for optimal cooling. In a thermal distribution the curve would be linear--clearly not the case here. (b) Power spectrum Eq.~\eqref{eq:spectrum5} of the motional sidebands for $\psi=\pi/2$ consisting of Lorentzians centered around the transition frequencies $\omega_{n,n+1}=(2n+3)\nu$. Transitions of the kind $n\rightarrow n\pm l$ with $l>1$ are not visible due to their small heights.}
\end{figure}

Figure~\ref{fig:well_doppler}(a) shows the dependence of the mean steady state occupation $\bar{m}=\Sigma_mp_mm$ on the laser detuning $\Delta$ for $\Omega=\Gamma/5$. 
%[E.g. for a linewidth $\Gamma=2\pi\times5.2\,{\rm MHz}$, corresponding to a D$_2$-line ($2\pi\times 351.7\,{\rm THz}$) in Cesium ($M=2.2\times 10^{-25}\,{\rm kg}$), these parameters result in a well length $L\approx37\,{\rm nm}$ and a Lamb-Dicke parameter $\eta\approx0.05$ ($kL\approx0.27$).]
Optimal Doppler cooling occurs at the detuning $\Delta\approx -0.59\Gamma$ (indicated by a circle), where a steady state occupation of $\bar{m}\approx1.24$ is reached. The optimal detuning takes a higher value than in the harmonic trapping potential where best Doppler cooling is achieved for approximately $\Delta=-\Gamma/2$ with $\bar{m}\approx 3$. This shows that in the square well potential the atomic motion is Doppler cooled well below the final value for harmonically trapped atoms. 

In the inset of Fig.~\ref{fig:well_doppler}(a) we plot the negative logarithm of the populations $p_m$ for optimal Doppler cooling against the eigenenergies $\varepsilon_m$. Since $-\log(p_m)$ does not show a linear dependence on $\varepsilon_m$ the steady state of the atomic motion cannot be written in the form of a thermal state, i.e. $\mu_{\rm st}\neq\exp(-\beta H_{\rm E})/Z$ with $H_{\rm E}=\sum_n (n+1)^2\hbar\nu \vert n\rangle\langle n\vert$.
 
 In Fig.~\ref{fig:well_doppler}(b) we show the motional sidebands detected under an angle $\psi=\pi/2$ calculated using Eq.~\eqref{eq:spectrum5} for optimal cooling $\Delta=-0.6\Gamma$ and a smallness parameter of $\eta=0.1$. The sideband spectrum consists of a series of Lorentzians centered around the transition frequencies between neighboring energy levels, viz. $\omega_{n,n+1}=(2n+3)\nu$ such that they are separated by $2\nu$. Their height decreases rapidly towards higher values of $n$, because of the lower populations $p_n$ of higher motional states. Transitions between non-adjacent levels are not visible due to the smallness of the position operator's matrix elements and vanishing values of $r(\omega)$. 
 
\subsubsection{Resolved sideband regime}
\begin{figure}[h!]
	\flushleft{(a)}\vspace{-4ex}
	\begin{center}
		\includegraphics[width=0.8\linewidth]{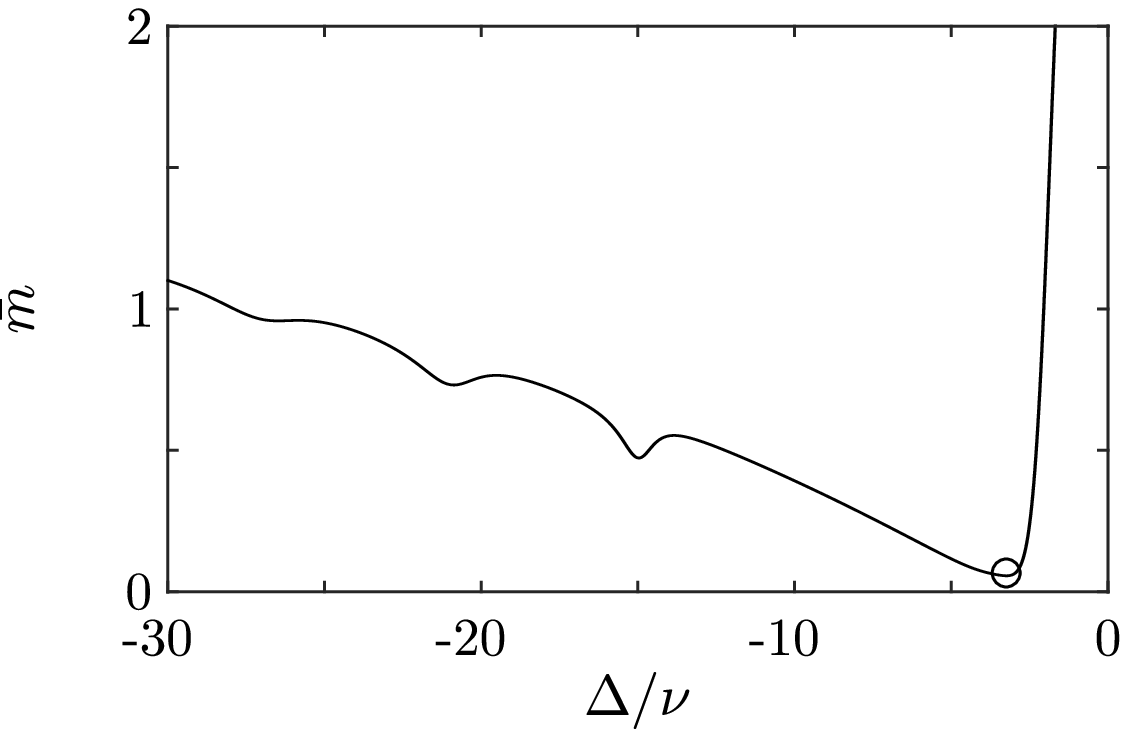}
	\end{center}
	\flushleft{(b)}\vspace{-4ex}
	\begin{center}
		\includegraphics[width=0.8\linewidth]{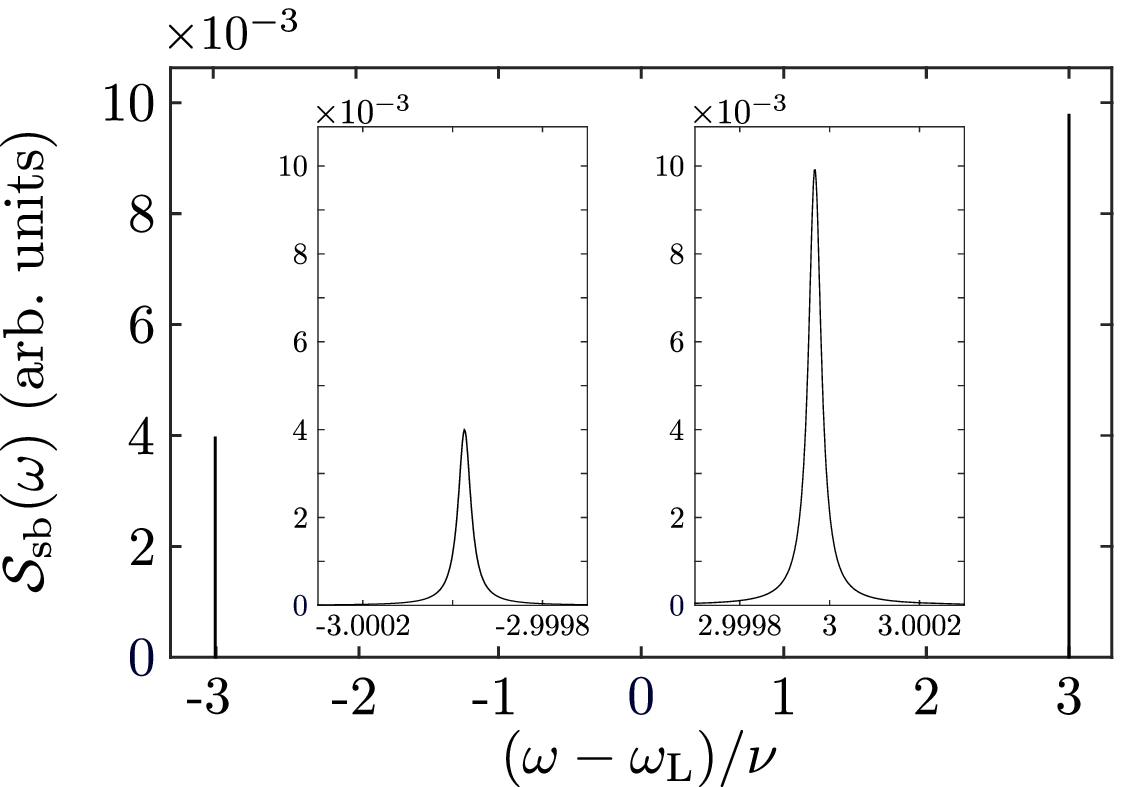}
	\end{center}
\caption{\label{fig:well_resolved}(a) Cooling of an atom trapped in an infinite square well potential in the resolved sideband regime for $\nu=10\Gamma/3$ (corresponding to $\omega_{01}=10\Gamma$) and $\Omega=\Gamma/5$. The mean occupation $\bar{m}$ in dependence on the laser detuning $\Delta$ takes its minimum $\bar{m}\approx0.086$ at $\Delta\approx -3.35\nu$ (indicated by a circle), where sideband cooling of the $1\rightarrow 0$ transition occurs. (b) Power spectrum of the motional sidebands for $\psi=\pi/2$ and $\eta=0.1$. The insets show magnifications of the first red and blue sideband which dominate the spectrum.}
\end{figure}

In the regime of resolved sideband cooling the transition frequency between the lowest energy states is much larger than the atomic linewidth. We choose $\nu=10\Gamma/3$, which corresponds to a transition frequency $\omega_{0,1}=10\Gamma$. Figure~\ref{fig:well_resolved}(a) shows again the dependence of the mean occupation on the laser detuning for $\Omega=\Gamma/5$.
Optimal cooling to a mean occupation $\bar{m}\approx0.068$ is achieved when the laser is red-detuned from the atomic transition by roughly the first transition frequency $\omega_{0,1}$, i.e $\Delta\approx-3.35\nu$. In contrast to Doppler cooling the cooling curve shows additional dips at the detunings $\Delta/\nu=-15,-21,-27,...$, corresponding to frequencies where multiple transition between motional eigenstates coincide, e.g. $\omega_{3,0}=\omega_{7,6}=-15\nu$ or $\omega_{4,1}=\omega_{10,9}=-21\nu$.

Due to ground state cooling, the spectrum shown in Fig.~\ref{fig:well_resolved}(b) is dominated by two peaks involving the lowest transition frequency $\pm\omega_{0,1}=\pm3\nu$. The inset shows a magnification of the red and blue sideband where also the frequency shift due to the interaction with the laser becomes visible. We also note that the half-width at half-maximum of both peaks is equal, due to the invariance of $\gamma_{nm}$ under exchange of $n$ and $m$, and that the asymmetry in the height vanishes when the spectrum is averaged over the detection angle $\psi$.

\subsection{Morse potential}
\label{subsec:morse}
\begin{figure}[ht]
	\begin{center}
		\includegraphics[width=0.7\linewidth]{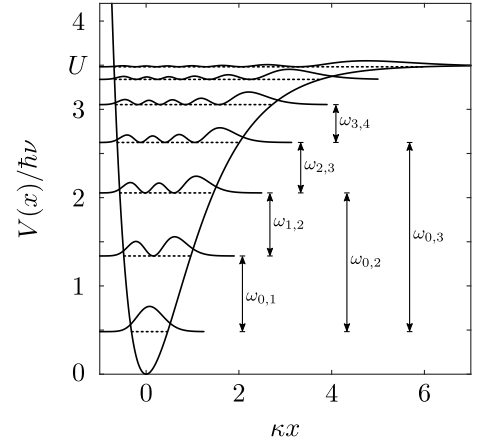}
		\caption{\label{fig:morse}The asymmetric Morse potential and the eigenenergies (dashed) and corresponding probability densities $|\varphi_n(x)|^2$ of its bound states for $a=7$ with some transition frequencies of possible transitions (in units of $\nu$).}
	\end{center}
\end{figure}

We turn to the discussion of the Morse potential~\cite{morse1929-1,morse1929-2} sketched in Fig.~\ref{fig:morse}. 
It has possible applications in the laser cooling of the vibrational degrees of freedom of di-atomic molecules~\cite{nguyen2011,manai2013,hamamda1015}. Although the general results are well-known, we briefly summarize them for the sake of completeness. 

The potential has the form
\begin{align}
\label{eq:morse_potential}
V(x)=U \left(1-e^{-\kappa x}\right)^2
\end{align}
with the depth $U$ and the characteristic length scale $1/\kappa$. Its eigenenergies are given by~\cite{pekeris1934}
\begin{align}
\label{eq:morse_energies}
\varepsilon_n=\left[\left(n+\frac{1}{2}\right)-\frac{1}{2a}\left(n+\frac{1}{2}\right)^2\right]\hbar\nu
\end{align}
for $n=0,1,2,...$ with the frequency $\nu=\kappa\sqrt{2U/M}$ and the dimensionless parameter $a=\sqrt{2MU}/\hbar\kappa$. The corresponding energy eigenstates $\vert n\rangle$ in the position representation read~\cite{dahl1988}
\begin{align}
\varphi_n(x)=&\langle x\vert n\rangle\nonumber\\
=&\mathcal{N}_n \zeta^{a-n-1/2}e^{-\zeta/2}L_n^{(2a-2n-1)}(\zeta)
\end{align}
with the abbreviation $\zeta=2a\exp(-\kappa x)$ and the generalized Laguerre polynomials $L_n^{(\alpha)}(z)$ and
$
\mathcal{N}_n=[\kappa(2a-2n-1)n!/\Gamma(2a-n)]^{1/2}
$~\cite{pekeris1934,scholz1932}.
The highest bound state in this potential has the index $n=\left \lfloor{a-1/2}\right \rfloor$. Figure~\ref{fig:morse} shows the Morse potential and its eigenenergies in units of $\hbar\nu$ along with the probability density of the bound states for $a=7$. The transition frequency between adjacent energy eigenstates decreases according to 
\begin{align}
\omega_{n,n+1}=\frac{a-1-n}{a}\nu
\end{align}

The matrix elements of the position operator required for the evaluation of the transition rates $A_{nm}$ were reported in~\cite{scholz1932,gallas1980,sandoval1990} and can be written in the form
\begin{align}
&\langle n\vert x\vert m\rangle=\frac{(-1)^{n+m}}{\kappa(n-m)(2a-n-m-1)}\nonumber\\
&\times\left[\frac{m!}{n!}\frac{\Gamma(2a-m)}{\Gamma(2a-n)}(2a-2n-1)(2a-2m-1)\right]^{1/2}
\end{align}
for $n<m$, while for $n>m$ the two indices have to be interchanged on the right-hand side. The moments of the position operator in the eigenstates of the Morse potential are obtained from the generating function~\cite{gallas1980} 
\begin{align}
\langle n\vert e^{sx}\vert n\rangle&=\frac{(2a-2n-1)n!}{\Gamma(2a-n)}e^{s\log(2a)/\kappa}\nonumber\\
&\times\sum_{j,l=0}^n\frac{(-1)^{j+l}}{j!l!}\binom{2a-2n-1}{n-j}\binom{2a-2n-1}{n-l}\nonumber\\
&\times \Gamma(2a-2n-1+j+l-s/\kappa)
\end{align}
by differentiation with respect to $s$. For the ground state $n=0$ the first and second moments are
\begin{align}
\langle 0\vert x\vert 0\rangle&=\frac{1}{\kappa}\big[\log(2a)-\psi^{(0)}(2a-1)\big],\\
\langle 0\vert x^2\vert 0\rangle&=\frac{1}{\kappa^2}\big[\psi^{(1)}(2a-1)+\langle 0\vert x\vert 0\rangle^2\big]
\end{align}
with the polygamma functions $\psi^{(n)}(z)$~\cite{abramowitz1965}, resulting in the ground-state variance $\xi^2=\psi^{(1)}(2a-1)/\kappa^2$ and thereby  a smallness parameter
\begin{align}
\label{eq:lamb-dicke-morse}
\eta=\frac{k}{\kappa}\sqrt{\psi^{(1)}(2a-1)}.
\end{align}

\subsubsection{Doppler regime}
\begin{figure}[htb!]
\flushleft{(a)}\vspace{-4ex}
\begin{center}
\includegraphics[width=0.8\linewidth]{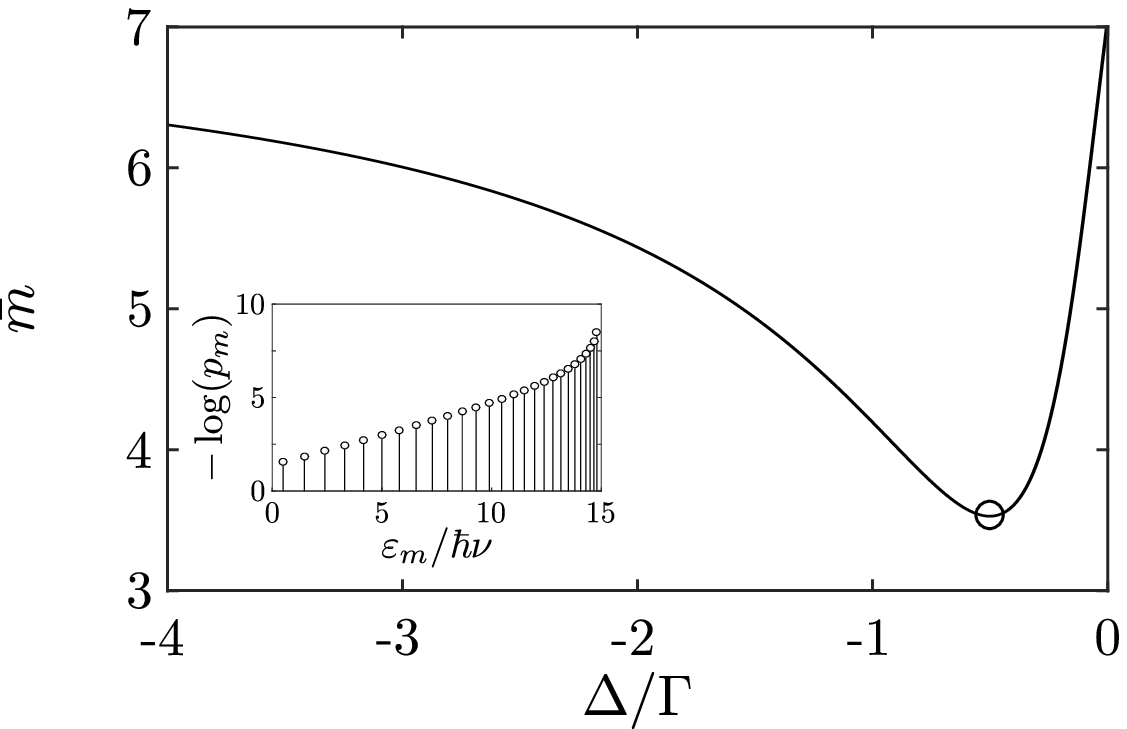}
\end{center}
\flushleft{(b)}\vspace{-4ex}
\begin{center}
\includegraphics[width=0.8\linewidth]{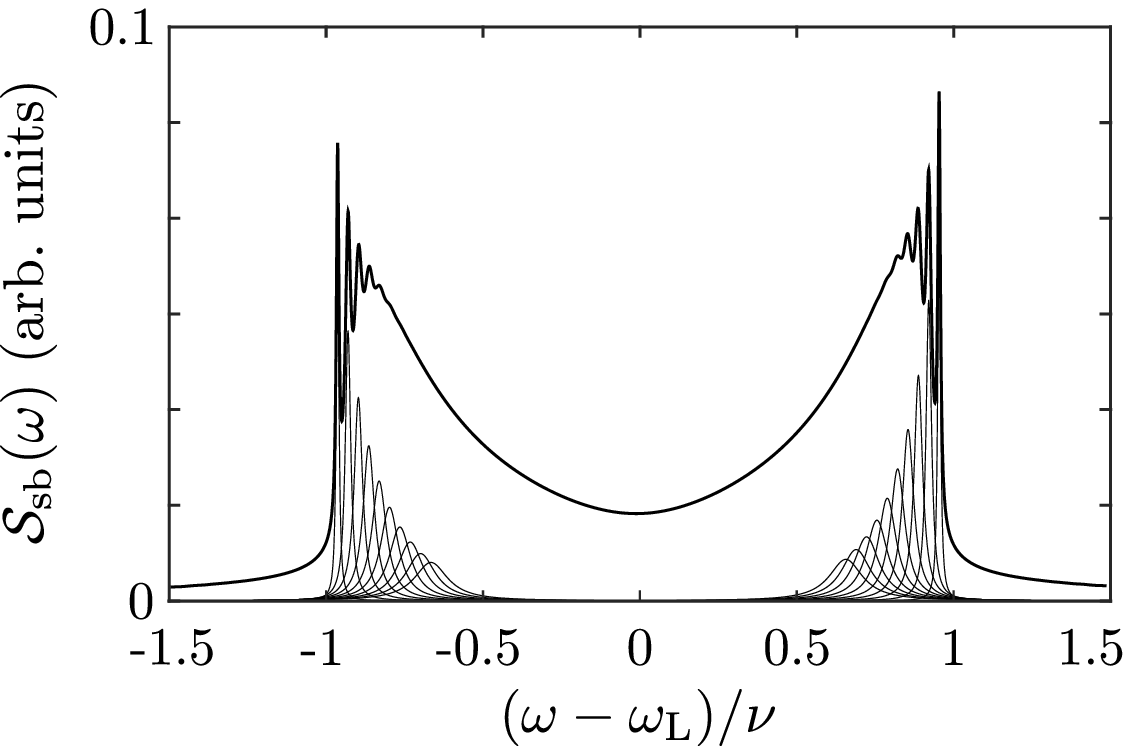}
\end{center}
\caption{\label{fig:morse_doppler}Cooling of atoms trapped in a Morse potential in the Doppler regime for $a=30$, $\nu=a\Gamma/10(a-1)$ (corresponding to $\omega_{01}=\Gamma/10$) and $\Omega=\Gamma/5$. (a) Mean occupation $\bar{m}$ in dependence of the laser detuning $\Delta$. The circle indicates optimal cooling occurring at $\Delta\approx-0.51\Gamma$ resulting in $\bar{m}=\Sigma_mmp_m\approx3.54$. The inset shows logarithm of the steady state occupations $p_m$ ($m=0,...,24$) for optimal cooling. In a thermal distribution the resulting curves would be linear--only approximately the case for the lower energy levels. (b) Power spectrum of the motional sidebands. The thin lines correspond to the first ten transitions between neighboring states on both sidebands. The smallness parameter is given by $\eta=0.1$ and the detector angle by $\psi=\pi/2$.}
\end{figure}

In Fig.~\ref{fig:morse_doppler} we show the mean steady state occupation $\bar{m}$ in dependence of $\Delta$ for $a=30$ and $\Omega=\Gamma/5$ assuming that the transition frequency between the lowest two states is again given by $\omega_{01}=\Gamma/10$ which leads to $\nu=a/(a-1) \Gamma/10\approx0.1034\Gamma$. Optimal cooling is achieved for $\Delta\approx0.509\Gamma$ with a mean occupation of $\bar{m}\approx3.54$.
In the inset we plot $-\log(p_m)$ against the eigenenergies $\varepsilon_m$. It can be seen that only the populations of the lower energy levels can be approximated by a thermal distribution while for higher states this is clearly not the case. In the calculation only the finite number $\left \lfloor{a-1/2}\right \rfloor+1$ of bound states of the potential, here 30 for $a=30$, are taken into account while free solutions are disregarded. This is a good approximation for states which are energetically well localized within the range of bound states.

In Fig.~\ref{fig:morse_doppler}(b) we show the motional sidebands of an atom trapped in a Morse potential ($a=30$) calculated using Eq.~\eqref{eq:spectrum5}. We used the parameters of optimal cooling, i.e. $\Delta=-0.509\Gamma$, a detection angle $\psi=\pi/2$ and a smallness parameter $\eta=0.1$. The decreasing transition frequency between neighboring states is reflected in the fact that in the sideband spectrum the modulus of the peak positions is smaller than the first transition frequency $\omega_{01}$. The thin curves under the sideband spectrum are the main components, according to the decomposition Eq.~\eqref{eq:spectrum5}, corresponding to the transitions $n+1\rightarrow n$ for $n=0,...,9$ on the blue sideband and $n-1\rightarrow n$ for $n=1,...,10$ on the red sideband, which add up to the complete spectrum but overlap because of their finite width.

\subsubsection{Resolved sideband regime}
In the resolved sideband case with $\nu=10a/(a-1)\Gamma$, corresponding to $\omega_{01}=10\Gamma$, optimal cooling occurs when the laser is red-detuned by the transition frequency $\omega_{01}$, leading to a mean occupation $\bar{m}=0.0026$. This is only slightly higher than in the harmonic case where $\bar{m}=0.0016$ is achieved (calculated using $\bar{m}=A_+/(A_--A_+)$, see Sec.~\ref{sec:harmonic_comparison}). The fact that in this case the cooling is more efficient than in the square well case is due to the comparatively small anharmonicity of the Morse potential in the lower energy levels.
\begin{figure}[htb!]
	\flushleft{(a)}\vspace{-4ex}
	\begin{center}
		\includegraphics[width=0.8\linewidth]{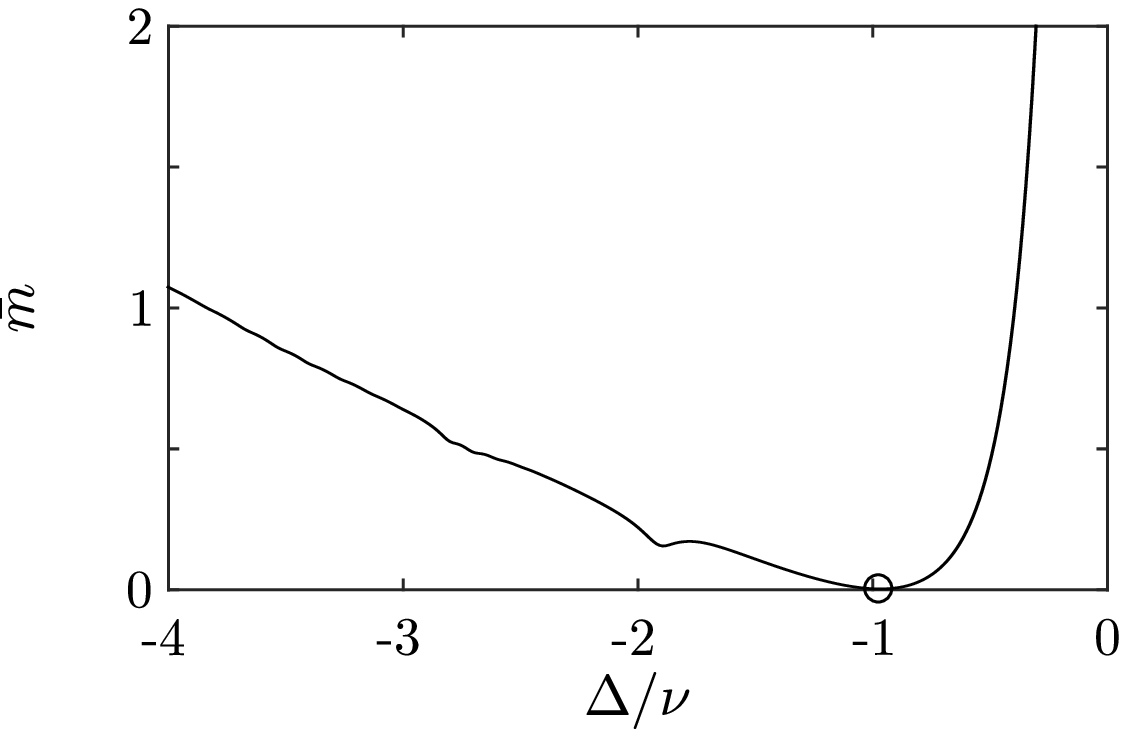}
	\end{center}
	\flushleft{(b)}\vspace{-4ex}
	\begin{center}
		\includegraphics[width=0.8\linewidth]{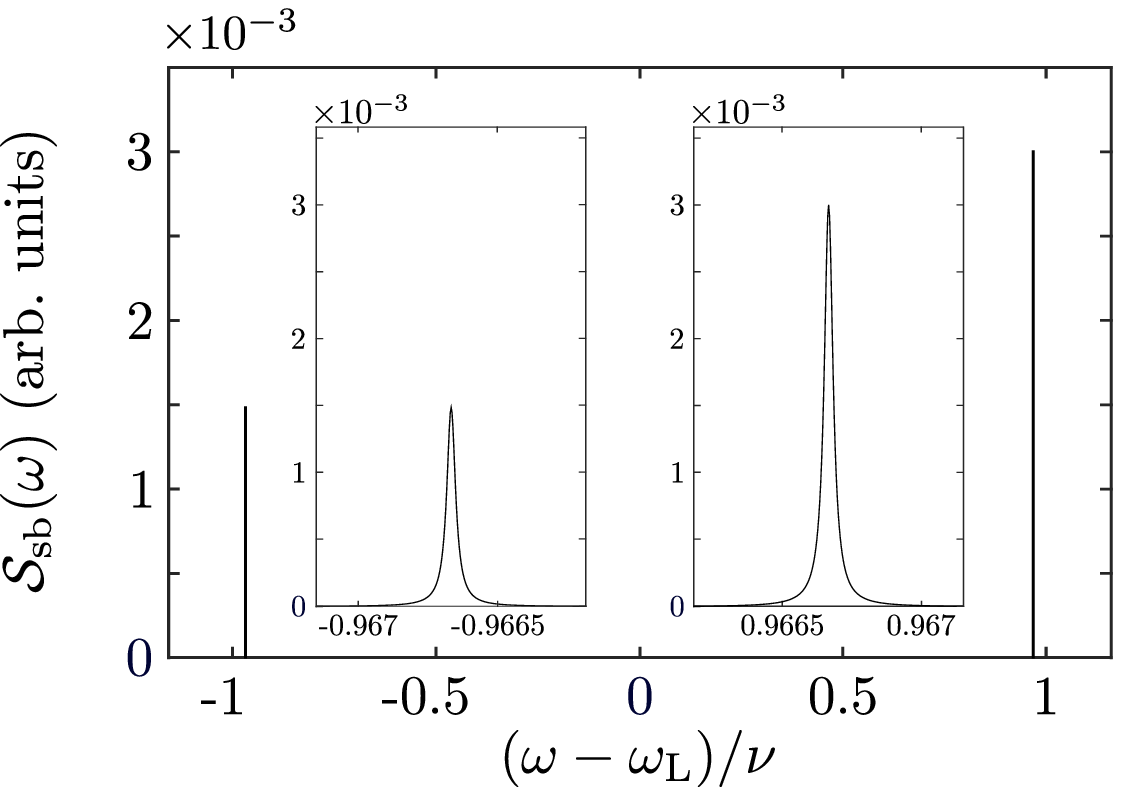}
	\end{center}
	\caption{\label{fig:morse_resolved}Cooling of an atom trapped in a Morse potential in the resolved sideband regime for $\nu=10a/(a-1)\Gamma$ (corresponding to $\omega_{01}=10\Gamma$) and $\Omega=\Gamma/5$. The mean occupation $\bar{m}=\Sigma_mmp_m$ in dependence on the laser detuning $\Delta$ takes its minimum $\bar{m}\approx0.0026$ at $\Delta\approx -\omega_{01}$ (indicated by a circle), where sideband cooling of the $1\rightarrow 0$ transition occurs. (b) Power spectrum of the motional sidebands for a smallness parameter $\eta=0.1$ and $\psi=\pi/2$.}
\end{figure}
 Figure~\ref{fig:morse_resolved}(a) shows the mean occupation in dependence of the detuning $\Delta$ where again shallow dips are visible where multiple transitions frequencies coincide, for example $\omega_{2,0}=\omega_{12,9}=-1.9\nu$, $\omega_{4,1}=\omega_{19,13}=-2.7\nu$, and $\omega_{3,0}=\omega_{21,14}=-2.8\nu$.

In this regime the motional sideband spectrum, depicted in Fig.~\ref{fig:morse_resolved}(b), is also mainly given by the two peaks arising from the transitions between the lowest energy states. The insets show a magnification of the red and blue sideband.

\section{Conclusion}
\label{sec:conclusion}
Following up on Dicke's original work on light scattered by atoms confined to an infinite square well and subsequent works on harmonically trapped atoms we present the spectrum of resonance fluorescence of laser cooled atoms trapped in arbitrary potentials. 
The treatment relies on a perturbative analysis of the power spectrum of the scattered light up to second order in the Lamb-Dicke parameter and is based on the solution of the corresponding master equation describing the laser cooling dynamics. 

We applied the results to two exemplary potentials, the infinite square well and the Morse potential, and distinguished between different cooling regimes, namely the Doppler and resolved sideband cooling. In contrast to the harmonic trapping potential the steady state of the atomic motion does not take the form of a thermal distribution. The spectrum of the motional sidebands consists of a series of peaks centered around the transitions between eigenstates of the atomic center-of-mass motion. From our treatment also follow the widths of the individual peaks which are determined by the transitions rates between the motional states due to the interaction with the laser.  In the case of separated motional sideband peaks, the temperature of the cooled atom can be extracted from the individual heights. When the peaks overlap, the resulting asymmetric envelope of the spectral signals stemming from the atomic motion can be used to determine the temperature by using our theoretical results.

Asymmetric motional sidebands have been observed in recent atom-cavity experiments where instead of the resonance fluorescence the spectrum of the cavity output was measured. In this case the atom is trapped in the sinusoidal optical-lattice potential of the cavity mode. With the work presented here we lay the ground work for further investigation of such periodic potentials offering possible means for temperature extraction in cavity-cooling experiments.

\section{Acknowledgments}
The authors gratefully acknowledge helpful discussions with Giovanna Morigi and financial support from the GradUS program of the Saarland University, the German Ministry of Education and Research (BMBF "Q.com") and the German Research Foundation (DFG) within the Project No.
BI1694/1-1.

\appendix
\section{Matrix representation of the internal Liouville operator}
\label{app:internal}
In the basis $\{\vert g\rangle\langle g\vert,\vert e\rangle\langle g\vert,\vert g\rangle\langle e\vert,\vert e\rangle\langle e\vert\}$ the internal Liouville operator in Eq.~\eqref{eq:liouvillians0} can be written in its matrix representation~\cite{jakob2003}
\begin{align}
\label{eq:internal_liouvillian}
\mathcal{L}_{\rm I}=\frac{1}{2}\begin{bmatrix}
0 & -i\Omega &  i\Omega & 2\Gamma\\
-i\Omega & 2i\tilde{\Delta} &  0 &  i\Omega\\
i\Omega &  0 &  -2i\tilde{\Delta}^\ast & -i\Omega\\
0 & i\Omega & -i\Omega & -2\Gamma
\end{bmatrix}
\end{align}
with the complex detuning $\tilde{\Delta}=\Delta+i\Gamma/2$. From this we can easily derive the actual form of the internal eigenelements, being the eigenvectors of the matrix~\eqref{eq:internal_liouvillian}. For instance, the steady state $\rho_{\rm st}$ of the internal dynamics, fulfilling $\mathcal{L}_{\rm I}\rho_{\rm st}=0$, is given by 
\begin{align}
\label{eq:internal_steady}
\rho_{\rm st}=\frac{1}{N} \begin{pmatrix}
|\tilde{\Delta}|^2+\Omega^2/4 &  \tilde{\Delta}\Omega/2\\
\tilde{\Delta}^\ast\Omega/2 & \Omega^2/4\\
\end{pmatrix}
\end{align}
in the basis $\{\vert g\rangle, \vert e\rangle\}$, with the normalization constant $N=\Gamma^2/4+\Delta^2+\Omega^2/2$ defined in the main text.

\section{Weight factors}
\label{app:sidebands}
We will outline the main steps in the evaluation of the weight factors that contribute to the motional sidebands. They can be calculated according to 
\begin{widetext}
\begin{align}
{\rm Tr}\big\{D_-^{(0)}\hat{\varrho}^{(1)}_\lambda+D_-^{(1)}\hat{\varrho}^{(0)}_\lambda\big\}=&{\rm Tr}\big\{\sigma_+\big[(\lambda_{nm}-\mathcal{L}_0)^{-1}\mathcal{Q}_{nm}\mathcal{L}_1+ik\cos\psi\,x\big]\rho_{\rm st}\hat{\mu}_{nm}\big\}\nonumber\\
=&-i\left[ r(\omega_{nm})-k\cos\psi {\rm Tr}\{\sigma_+\rho_{\rm st}\}\right]{\rm Tr}\{x\hat{\mu}_{nm}\},\\
{\rm Tr}\big\{\big(\check{\varrho}^{\dagger(1)}_\lambda D_+^{(0)}+\check{\varrho}^{\dagger(0)}_{\lambda}D_+^{(1)}\big)\varrho^{(0)}_{\rm st}\big\}
=&{\rm Tr}\big\{\check{\mu}_{nm}^\dag\big[\mathcal{L}_1(\lambda_{nm}-\mathcal{L}_0)^{-1}\mathcal{Q}_{nm}-ik\cos\psi\,x\big]\sigma_+\rho_{\rm st}\hat{\mu}_{nm}\big\}\nonumber\\
=-i\Big[\frac{1}{\hbar}{\rm Tr}\big\{W_1&(\lambda_{nm}-\mathcal{L}_{\rm I})^{-1}\mathcal{Q}_{nm}\sigma_+\rho_{\rm st}\big\}{\rm Tr}\{\check{\mu}^\dag_{nm}[x,\mu_{\rm st}]\}+k\cos\psi {\rm Tr}\{\sigma_+\rho_{\rm st}\}{\rm Tr}\{\check{\mu}^\dag_{nm}x\}\Big]\\
{\rm Tr}\big\{\check{\varrho}^{\dagger(0)}_{\lambda}D_+^{(0)}\varrho^{(1)}_{\rm st}\big\}=&-{\rm Tr}\big\{\check{\mu}_{nm}^\dag\sigma_+\mathcal{L}_0^{-1}\mathcal{Q}_0\mathcal{L}_1\rho_{\rm st}\mu_{\rm st}\big\}\nonumber\\
=i\Big[r^\ast(\omega_{nm})&{\rm Tr}\{\check{\mu}^\dag_{nm}x\mu_{\rm st}\}+\frac{1}{\hbar}{\rm Tr}\big\{\sigma_+(\lambda_{nm}+\mathcal{L}_{\rm I})^{-1}\mathcal{Q}_{nm}\rho_{\rm st}W_1\big\}{\rm Tr}\{\check{\mu}^\dag_{nm}[x,\mu_{\rm st}]\}\Big].
\end{align}
\end{widetext}
where we have used the corrections Eq.~\eqref{eq:first_order_correction1}, Eq.~\eqref{eq:first_order_correction2}, Eq.~\eqref{eq:gen_dipole1} and Eq.~\eqref{eq:gen_dipole2} as well as the explicit form of $\mathcal{L}_1$ and the projector $\mathcal{Q}_{nm}=1-\mathcal{P}_{\lambda_{nm}}$. Addition of these terms yields the result
\begin{align}
\label{app:eq:9}	
w^{(2)}_{\lambda}&=\vert\langle n\vert x\vert m\rangle\vert^2\Big[p_m\big| r(\omega_{nm})-k\cos\psi {\rm Tr}\{\sigma_+\rho_{\rm st}\}\big|^2 \nonumber\\
+&(p_n-p_m)\big(r(\omega_{nm})-k\cos\psi {\rm Tr}\{\sigma_+\rho_{\rm st}\}\big)q(\omega_{nm})\Big],
\end{align}
where we already inserted the actual form of the motional steady state $\mu_{\rm st}=\Sigma_j p_j \vert j\rangle\langle j\vert$ and the two definitions Eqs.~\eqref{eq:r} and~\eqref{eq:q}. By evaluating ${\rm Tr}\{\sigma_+\rho_{\rm st}\}=[\Delta+i\Gamma/2]\Omega/2N$, with help of the internal steady state~\eqref{eq:internal_steady}, this corresponds to Eq.~\eqref{eq:spectrum4}.

\section{Explicit expressions for the functions $r(\omega)$, $q(\omega)$ and $s(\omega)$}
\label{app:r_and_q}
In this section we give explicit expressions for the two functions $r(\omega)$ and $q(\omega)$ from Eqs.~\eqref{eq:r} and~\eqref{eq:q} as well as $s(\omega)$ from Eq.~\eqref{eq:s_omega}. Using the quantum regression theorem~\cite{carmichael2002} $\langle X(t)Y(0)\rangle_{\rm st}={\rm Tr}\{X\exp(\mathcal{L}t)Y\varrho_{\rm st}\}$ and formal integration $\int_0^\infty {\rm d}t\, \exp(-z+\mathcal{L})t=(z-\mathcal{L})^{-1}$ they can be rewritten as
\begin{widetext}
\begin{align}
r(\omega)=&\frac{1}{\hbar}{\rm Tr}\big\{\sigma_+(i\omega-\mathcal{L}_{\rm I})^{-1}[W_1,\rho_{\rm st}]\big\}=\frac{\Omega}{2}k\cos\phi\frac{\Gamma N \tilde{\Delta}+i[(\tilde{\Delta}+i\Gamma)|\tilde{\Delta}|^2+\Delta\Omega^2]\omega-i|\tilde{\Delta}|^2\omega^2}{\Gamma N^2+i[5\Gamma^2/4+\Delta^2+\Omega^2]N\omega-2\Gamma N\omega^2-iN\omega^3},\\
q(\omega)=&\frac{1}{\hbar}{\rm Tr}\big\{W_1(i\omega-\mathcal{L}_{\rm I})^{-1}\sigma_-\rho_{\rm st}\big\}-\frac{1}{\hbar}{\rm Tr}\big\{\sigma_-(i\omega+\mathcal{L}_{\rm I})^{-1}\rho_{\rm st}W_1\big\}\nonumber\\
=&\frac{\Omega^3}{2}k\cos\phi\frac{i\Gamma^3 N\tilde{\Delta}^\ast-\Gamma^2\tilde{\Delta}^\ast N\omega+[\Gamma^2(B+\Omega^2)-i\Gamma B\tilde{\Delta}^\ast]\omega^2+B\tilde{\Delta}^\ast\omega^3+[B-i\Gamma\tilde{\Delta}]\omega^4+\tilde{\Delta}^\ast\omega^5+\omega^6}{2\Gamma^2N^3\omega+[\Gamma^2(9\Gamma^2/8-3\Delta^2+\Omega^2)+2(\Delta^2+\Omega^2)^2]N\omega^3+4BN\omega^5},\\
s(\omega)=&-\frac{1}{\hbar^2}{\rm Tr}\big\{W_1(i\omega+\mathcal{L}_{\rm I})^{-1}W_1\rho_{\rm st}\big\}\nonumber\\
=&\frac{\Omega^2}{4}k^2\cos^2\phi \frac{\Gamma^3\Omega^2/(4N)+[\Gamma (\Delta-3i\Gamma/2)+i\Gamma^2|\tilde{\Delta}|^2/N]\omega+[i\Delta|\tilde{\Delta}|^2/N-3\Gamma/2-2i\Delta]\omega^2+i\omega^3}{-i\Gamma N\omega-[5\Gamma^2/4+\Delta^2+\Omega^2]\omega^2+2i\Gamma \omega^3+\omega^4}.
\end{align}
\end{widetext}
with $\tilde{\Delta}=\Delta+i\Gamma/2$ and the constant $B=3\Gamma^2/4-\Delta^2-\Omega^2$. In the evaluation of the trace we employed the explicit form of the matrices $(i\omega\mp\mathcal{L}_{\rm I})^{-1}$ which can be obtained by using the Liouville operator Eq.~\eqref{eq:internal_liouvillian}.

\section{Second order eigenvalue corrections}
\label{app:sidebands_width}
We briefly go into the derivation of the perturbative corrections of the eigenvalues that lead to a non-vanishing real part which becomes immanent in the finite width of the sideband peaks. Their form is given by
\begin{align}
\label{app:eq:12}	
\lambda_2=&{\rm Tr}\big\{\check{\varrho}^{\dagger(0)}_\lambda\big[\mathcal{L}_2+\mathcal{L}_1(\lambda_0-\mathcal{L}_0)^{-1}\mathcal{Q}_\lambda\mathcal{L}_1\big]\hat{\varrho}^{(0)}_\lambda\big\}.
\end{align}
We will treat the two terms separately, starting with the first one that involves $\mathcal{L}_2$. For $\lambda_0=\lambda_{nm}$ this part yields
\begin{align}
\label{app:eq:13}	
{\rm Tr}&\{\check{\mu}^\dag_{nm}\mathcal{L}_2\rho_{\rm st}\hat{\mu}_{nm}\}=\frac{1}{i\hbar}{\rm Tr}\{W_1\rho_{\rm st}\}\big[\langle n\vert x^2\vert n\rangle-\langle m\vert x^2\vert m\rangle\big]\nonumber\\
&+\frac{D}{2}\big[2\langle n\vert x\vert n\rangle \langle m\vert x\vert m\rangle-\langle n\vert x^2\vert n\rangle-\langle m\vert x^2\vert m\rangle\big]
\end{align}
The second term in~\eqref{app:eq:13} can be brought in the form
\begin{align}
\label{app:eq:14}	
{\rm Tr}&\big\{\check{\mu}^\dag_{nm}\mathcal{L}_1(\lambda_{nm}-\mathcal{L}_0)^{-1}\mathcal{Q}_\lambda\mathcal{L}_1\rho_{\rm st}\hat{\mu}_{nm}\big\}=\nonumber\\
&-\sum_j \left[s^\ast(\omega_{jn})\vert\langle m\vert x\vert j\rangle\vert^2+s(\omega_{jm})\vert\langle n\vert x\vert j\rangle\vert^2\right].
\end{align}
After recombining these two results using $\langle n\vert x^2\vert n\rangle=\Sigma_j \vert\langle n\vert x\vert j\rangle\vert^2$ we obtain
\begin{align}
\label{eq:eigenvals4}
\lambda_2^{nm}=&\frac{i\Delta\Omega^2k^2\cos^2\phi}{4N} \left(\langle n\vert x^2\vert n\rangle-\langle m\vert x^2\vert m\rangle\right)\nonumber\\
&-\frac{1}{2}\sum_j\left[2s(\omega_{jn})+D\right] \vert\langle j\vert x\vert n\rangle\vert^2\nonumber\\
&-\frac{1}{2}\sum_j\left[2s^\ast(\omega_{jm})+D\right] \vert\langle j\vert x\vert m\rangle\vert^2\nonumber\\
&+D\langle n\vert x\vert n\rangle \langle m\vert x\vert m\rangle.
\end{align}
With the definition of the transition rates $A_{nm}$ given by Eq.~\eqref{eq:matrix_coefficients1} this corresponds to Eqs.~\eqref{eq:eigenvals3}-~\eqref{eq:linewidths}.

\section{Rate equation}
\label{app:external_steady}
We outline the main steps to transform the equation of the reduced dynamics in the subspace for $\lambda=0$, i.e. Eq.~\eqref{eq:reduced_dynamics}, into a linear system $\mathcal{A}\mathbf{p}=0$ with $\mathbf{p}=(p_0,p_1,...)^T$. Since we aim at determining the steady state of the external dynamics it is necessary to trace out the internal degrees of freedom in~\eqref{eq:reduced_dynamics} yielding
\begin{align}
\label{app:eq_1}	
{\rm Tr}_{\rm I}\big\{\mathcal{P}_0\left(\mathcal{L}_1\mathcal{L}_0^{-1}\mathcal{Q}_0\mathcal{L}_1-\mathcal{L}_2\right)\mathcal{P}_0\rho_{\rm st}\mu_{\rm st}\big\}=0.
\end{align}
The action of the projector $\mathcal{P}_0$ is given by 
\begin{align}
\mathcal{P}_0 \rho\mu=\rho_{\rm st}\sum_n\langle n\vert \mu\vert n\rangle \vert n\rangle\!\langle n\vert
\end{align}
with internal and external density operators $\rho$ and $\mu$, respectively.
For the second term in Eq.~\eqref{app:eq_1} we obtain
\begin{align}
\label{app:eq_2}	
\langle n\vert {\rm Tr}_{\rm I}\{\mathcal{L}_2\rho_{\rm st}\mu_{\rm st}\}\vert n\rangle
=D\sum_{m} (p_m-p_n) \vert \langle n\vert x\vert m\rangle\vert^2.
\end{align}
The evaluation of the first term is somewhat more involved and reads
\begin{align}
\label{app:eq_3}	
\langle n&\vert {\rm Tr}_{\rm I}\{\mathcal{L}_1\mathcal{L}_0^{-1}\mathcal{Q}_0\mathcal{L}_1\rho_{\rm st}\mu_{\rm st}\}\vert n\rangle\nonumber\\ 
&=\int_0^\infty dt\,\langle n\vert {\rm Tr}_{\rm I}\{\mathcal{L}_1e^{\mathcal{L}_0t}\mathcal{Q}_0\mathcal{L}_1\rho_{\rm st}\mu_{\rm st}\}\vert n\rangle\nonumber\\
&=2\,{\rm Re}\sum_{m} \big[p_ms(\omega_{n,m})-p_ns(\omega_{m,n})\big]\vert\langle n\vert x\vert m\rangle\vert^2	
\end{align}
with $s(\omega)$ defined in Eq.~\eqref{eq:s_omega}. Here we inserted the actual form of the Liouville operators and separated the internal from the external expressions. We can now recombine the two terms leading to
\begin{align}
\label{app:eq_5}	
\sum_{m}  A_{nm}p_m-\sum_{m} A_{mn}p_n=0.
\end{align}
This is the set of equations that determines $\mathbf{p}$ and can be expressed in the matrix notation  $\mathcal{A}\mathbf{p}=0$ with
\begin{align}
\label{eq:rateequationmatrix}	
\mathcal{A}=\begin{bmatrix}
-\!\sum\limits_{m\neq 0} A_{m0}& A_{01} & A_{02} & \dots\\
A_{10} & -\!\sum\limits_{m\neq 1} A_{m1} & A_{12} &\\
A_{20} & A_{21} & -\!\sum\limits_{m\neq 2} A_{m2} &\\
\vdots& & &\ddots 
\end{bmatrix}.
\end{align}

\bibliography{article}

\end{document}